\documentclass[conference]{IEEEtran}
\IEEEoverridecommandlockouts
\usepackage{cite}
\usepackage{amsmath,amssymb,amsfonts}
\usepackage{algorithmic}
\usepackage{graphicx}
\usepackage{textcomp}
\usepackage{xcolor, colortbl}
\usepackage{booktabs}
\usepackage{multirow}
\usepackage{xspace}
\usepackage{caption}
\usepackage{subcaption}
\usepackage{footnote}
\usepackage{url}
\usepackage{tcolorbox}
\usepackage{booktabs}
\usepackage{enumitem}
\usepackage{fixltx2e}
\usepackage{multicol}
\usepackage{balance}
\usepackage{listings}
\usepackage[normalem]{ulem}

\definecolor{Gray}{gray}{0.9}
\definecolor{codegreen}{rgb}{0,0.6,0}
\definecolor{codegray}{rgb}{0.73,0.38,0.06}
\definecolor{codepurple}{rgb}{0.70,0.27,0}
\definecolor{codemagenta}{rgb}{0.74,0.09,0.42}
\definecolor{codeoutput}{rgb}{0.5,0,0}
\definecolor{backcolour}{rgb}{0.96,0.96,0.96}

\def\BibTeX{{\rm B\kern-.05em{\sc i\kern-.025em b}\kern-.08em
    T\kern-.1667em\lower.7ex\hbox{E}\kern-.125emX}}

\begin{document}

\title{An Empirical Study on Code Comment Completion}



\author{
	\IEEEauthorblockN{Antonio Mastropaolo,
		Emad Aghajani,
    	Luca Pascarella,
		Gabriele Bavota
	}
	
    \textit{SEART @ Software Institute, Universit\`{a} della Svizzera italiana (USI), Switzerland}
}

\newboolean{showcomments}

\setboolean{showcomments}{true}

\ifthenelse{\boolean{showcomments}}
  {\newcommand{\nb}[2]{
    \fbox{\bfseries\sffamily\scriptsize#1}
    {\sf\small$\blacktriangleright$\textit{#2}$\blacktriangleleft$}
   }
  }
  {\newcommand{\nb}[2]{}
  }

\newcommand\TODO[1]{\textcolor{red}{\nb{TODO}{#1}}}
\newcommand\ANTONIO[1]{\textcolor{blue}{\nb{ANTONIO}{#1}}}
\newcommand\LUCA[1]{\textcolor{green}{\nb{LUCA}{#1}}}
\newcommand\EMAD[1]{\textcolor{violet}{\nb{EMAD}{#1}}}
\newcommand\GABRIELE[1]{\textcolor{red}{\nb{GABRIELE}{#1}}}

\newcommand{\ie}{\emph{i.e.,}\xspace}
\newcommand{\eg}{\emph{e.g.,}\xspace}
\newcommand{\etc}{etc.\xspace}
\newcommand{\etal}{\emph{et~al.}\xspace}
\newcommand{\secref}[1]{Section~\ref{#1}\xspace}
\newcommand{\figref}[1]{Fig.~\ref{#1}\xspace}
\newcommand{\listref}[1]{Listing~\ref{#1}\xspace}
\newcommand{\tabref}[1]{Table~\ref{#1}\xspace}
\newcommand{\tool}[1]{{\sc #1}\xspace}

\newcommand{\totalmethods}{497,328\xspace}

\maketitle


\begin{abstract}
Code comments play a prominent role in program comprehension activities. However, source code is not always documented and code and comments not always co-evolve. To deal with these issues, researchers have proposed techniques to automatically generate comments documenting a given code at hand. The most recent works in the area applied deep learning (DL) techniques to support such a task. Despite the achieved advances, the empirical evaluations of these approaches show that they are still far from a performance level that would make them valuable for developers.
We tackle a simpler and related problem: Code comment completion. Instead of generating a comment for a given code from scratch, we investigate the extent to which state-of-the-art techniques can help developers in writing comments faster. We present a large-scale study in which we empirically assess how a simple $n$-gram model and the recently proposed Text-To-Text Transfer Transformer (T5) architecture can perform in autocompleting a code comment the developer is typing. The achieved results show the superiority of the T5 model, despite the $n$-gram model being a competitive solution.
\end{abstract}



\section{Introduction}
\label{sec:intro}
Code comprehension can take up to 58\% of developers' time \cite{Xia:tse2018}. In such a process, code comments play a pivotal role \cite{deSouza:2005} helping developers in understanding the source code at hand. However, as shown in recent studies, code comments may completely lack \cite{Spinellis:IE} or not co-evolve with the related code \cite{Fluri:WCRE07,Fluri:SQJ09,Linares:ASE15,Wen:icpc2019}, even becoming misleading for developers.

To support developers in code comprehension activities, researchers have proposed techniques and tools aimed at automatically documenting a given code at hand. These approaches can be roughly classified into two categories: extractive \cite{Rodeghero:icse17,Haiduc:wcre2010,Sridhara:icse2011,Moreno:icpc2013} and abstractive \cite{Sridhara:icse2011,McBurney:tse2016,Hu:icpc2018,iyer:acl,Allamanis:2016,Hu:emse2020,haque:2020}. The former create a summary of a code component which includes information extracted from the component being summarized (\eg a class is summarized by exploiting a set of predefined templates filled in with information extracted from the class code). While simple to implement and efficient in terms of execution time, these approaches fall short in case the code component uses a poor vocabulary. Abstractive approaches try to overcome this limitation by including in the generated summaries information that is not present in the code component to document. Among those techniques, the ones using deep learning (DL) to generate natural language descriptions for a given snippet of code are on the rise \cite{Hu:icpc2018,iyer:acl,Allamanis:2016,Hu:emse2020,haque:2020}, with attempts also made for automatically updating comments given a code change \cite{Liu:ase2020}.

Despite the substantial improvements brought by DL techniques in addressing the \emph{code comment generation} problem, the findings reported even in the most recent empirical studies show how these techniques are still far from being useful tools for software developers. 

For example, in the recent work by Mastropaolo \etal \cite{Matropaolo:icse2021}, the authors show that state-of-the-art techniques are able to generate comments equivalent to those written by humans in only $\sim$10\% of cases. For this reason, we believe that completely relying on ``\emph{machines}'' to write comments is, as of today, a far-fetched goal that, while worth investigating, is unlikely to result in major breakthrough in the short term. In this work, we tackle the simpler problem of \emph{code comment completion}, in which the ``machine'' is in charge of completing a comment that the developer starts writing, similarly to what done for code tokens by code completion techniques \cite{Bruch:fse2009,Raychev:pldi,svyatkovskiy2020fast,brody2020neural}. 

The code comment completion problem has been firstly tackled by Ciurumelea \etal \cite{saner2020} in the context of Python code: They study whether a deep learning model can predict the next word that a developer is likely to type while commenting code. This is, to the best of our knowledge, the only work done in this area. Stemming from their idea, we present in this paper a large-scale study assessing the ability of a simple $n$-gram model and of the recently proposed Text-To-Text Transfer Transformer (T5) architecture \cite{raffel2019exploring} in supporting code comment completion for Java programs. As compared to the work by Ciurumelea \etal \cite{saner2020}, besides focusing on a different context (\ie Python \emph{vs} Java) we: (i) investigate the actual advantages brought by a DL-based model (T5) over a simpler $n$-gram model that can be trained in a fraction of the time required by T5; (ii) do not limit our study to predicting the single next word the developer is likely to type, but evaluate how the investigated techniques perform when asked to predict longer word sequences (\eg the next 10 words), providing a more advanced completion support to developers. We also study the complementarity of the two techniques and report qualitative examples of correct and wrong predictions to understand their strengths and limitations.

Our study has been run on a dataset composed by \totalmethods Java methods with their related comments. The achieved results can be summarized as follows. First, the T5 model outperforms the $n$-gram model, achieving superior performance in all the comment completion scenario we tested. Second, despite being more performant, the T5 model exploits as input not only the first part of the comment already written by the developer (also used by the $n$-gram model), but also a \emph{context} representing the relevant code for the comment to complete. This means that the T5 model, as we tested it, can only be used when the developer writes the comment after the code has been already implemented (the assumption made by approaches for automated code documentation \cite{Hu:icpc2018,iyer:acl,Allamanis:2016,Hu:emse2020,haque:2020}). Thus, the applicability of the $n$-gram model is higher (\ie it can be used also when the code is not yet implemented).


\section{T5 to Support Code Comment Completion} \label{sec:approach}

\newcommand{\ICTask}{\textit{Inner-comment\textsubscript{task}}\xspace}
\newcommand{\JTask}{\textit{Javadoc\textsubscript{task}}\xspace}
\newcommand{\IComment}{\textit{C\textsubscript{I}}\xspace}
\newcommand{\JDComment}{\textit{C\textsubscript{JD}}\xspace}
\newcommand{\FTJInstance}{\textit{FT\textsubscript{JD-sample}}\xspace}
\newcommand{\BLDataset}{\textit{L\textsubscript{dataset}}\xspace}
\newcommand{\BothDataset}{\textit{L\textsubscript{dataset}+JD\textsubscript{dataset}}\xspace}


The T5 model has been introduced by Raffel \etal \cite{raffel2019exploring} to support multitask learning in Natural Language Processing (NLP). The idea is to reframe NLP tasks in a unified text-to-text format in which the input and output are always text strings. For example, a single model can be trained to translate across languages and to autocomplete sentences. This is possible since both tasks can be represented in a text-to-text format (\eg in the case of translation, the input is a sentence in a given language, while the output is the translated sentence). The T5 is trained in two phases: \textit{pre-training}, which allows defining a shared knowledge-base useful for a large class of sequence-to-sequence tasks (\eg guessing masked words in English sentences to learn about the language), and \textit{fine-tuning}, which specializes the model on a specific downstream task (\eg learning the translation of sentences from English to German). We briefly overview the T5 model and explain how we adapted it for supporting code comment completion. Finally, we describe the hyperparameter tuning of the model and the decoding strategy for generating the predictions.

\subsection{An Overview of T5}
The T5 is based on the transformer model architecture that allows handling a variable-sized input using stacks of self-attention layers. When an input sequence is provided, it is mapped into a sequence of embeddings passed into the encoder. 
Raffel \etal \cite{raffel2019exploring} propose five variants of T5: \textit{\textit{small}}, \textit{\textit{base}}, \textit{\textit{large}}, \textit{\textit{3 Billion}}, and \textit{\textit{11 Billion}}. These variants differ in terms of architectural complexity, with the smaller model having 60M parameters against the 11B of the largest one. While the accuracy of the most complex variants is higher as compared to the less complex models, the training complexity increases with the number of parameters \cite{raffel2019exploring}. Considering our computational resources, we decided to use the simplest T5\textsubscript{\textit{small}} model and, for this reason, we expect the achieved results to be a lower bound for the performance of a T5-based model in the task of code comment completion.

\textbf{T5\textsubscript{\textit{small}} architectural details.}
The T5\textsubscript{\textit{small}} architecture is characterized by six blocks for encoders and decoders. The feed-forward networks in each block consist of a dense layer with an output dimensionality ($d_{ff}$) of 2,048. The \textit{key} and \textit{value} matrices of all attention mechanisms have an inner dimensionality ($d_{kv}$) of 64, and all attention mechanisms have eight heads. All the other sub-layers and embeddings have a dimensionality ($d_{model}$) of 512.

\subsection{Problem Definition}\label{sub:problem}
We instantiate the T5 to the problem of code comment completion in Java. We tackle the problem at method-level granularity, meaning that we expect the model to learn how to autocomplete a code comment used to document a method or part of it. In Java, a method can be documented using a Javadoc comment (that we indicate with \JDComment) or inner comments (\IComment). Each comment is relevant to a specific \emph{context}. For example, the context of a \JDComment is the entire method, while the context of an \IComment can be a single line or a code block.

Given a \emph{context} and an incomplete comment (either a \JDComment or a \IComment), the trained model must predict the tokens needed to complete the comment. This implies that we must build a training dataset in which code comments are linked to the relevant part of the code they document (\ie the context). While this is trivial for \JDComment comments, a heuristic is needed for \IComment comments, since they are not explicitly linked to certain statements. \secref{dataset-prep} describes how we built such a dataset.

We pre-train the T5 by randomly masking tokens in comments asking the model to guess the masked tokens (\secref{subsec:training_strategy}). This builds the shared knowledge that we then specialize in two fine-tuning tasks, namely  \ICTask and \JTask, consisting in predicting the missing part of inner and Javadoc comments, respectively (\secref{subsec:finetuning_strategy}). 

\subsection{Dataset Preparation}\label{dataset-prep}
We start from the \emph{CodeSearchNet} dataset \cite{husain2019codesearchnet}, providing 6M functions from open-source projects. We only focus on the Java subset, composed of $\sim$1.5M methods.
We extract the set of instances using the {\tt docstring} (\ie the method's \emph{Javadoc}) and {\tt function} fields. Then, we run a pre-processing aimed at preparing our dataset. 

First, we discarded instances having \textit{\#tokens} $\geq$ 256, where \textit{\#tokens} = \textit{\#function\_tokens} +  \textit{\#docstring\_tokens} (\ie \textit{\#tokens} is the total number of tokens used to represent both the method and the comments associated to it). Such a filter is needed to limit the computational expense of training the model, and removed $\sim$9\% of instances from the dataset. Then, we excluded instances containing non ASCII characters as well as comments composed by less than three tokens (words), since unlikely to represent an interesting scenario for code comment completion. We also excluded comments representing instances of self-admitted technical debt (SATD) \cite{Potdar:icsme2014} (\ie comments documenting temporary workaround). Such a choice was dictated by the fact that we are interested in training our model to complete comments describing a method (or part of it) rather than comments used to document technical debt and very likely to be project-specific. We adopted a simple heuristic to discard SATD comments, excluding all comments starting with \texttt{TOFIX}, \texttt{TODO}, and \texttt{FIXME}. We are aware that more complex state-of-the-art techniques for SATD detection could be used (see \eg \cite{Ren:tosem2019}), but we preferred a simpler unsupervised heuristic for our pre-processing pipeline. 

We discarded commented code statements using the \emph{codetype} library \cite{codetype}. Then, we cleaned comments by replacing links with a special \textit{\_LINK\_} token (using the \emph{urlextract} library \cite{urlextract}), dates with a \textit{\_NUM\_} token (using the \emph{datefinder} library \cite{datefinder}), and references to code components with a \textit{\_REF\_} token. The latter are only handled in Javadoc comments exploiting the \texttt{@link} tag used to reference code elements. We further clean Javadoc comments by stripping \emph{HTML/XML} tags using the \emph{BeautifulSoup} library \cite{bsoup}. 

Then, we removed from each method all \IComment (inner comments) that are ``\textit{orphans}'', \ie \IComment followed \textbf{and} preceded by at least one blank line. As previously explained, to train the T5, we need to link each comment to its \emph{context}. 

Javadoc comments are linked to the whole method, while for inner comments we adopt a heuristic that would not work with orphan comments (\ie we cannot know what lines of code they likely document) --- details in \secref{subsec:finetuning_strategy}. 
As a last step in the processing of inner comments, we merge in a single \IComment subsequent inline comments that are not interleaved by empty lines or code statements. This is done since they likely represent a single multi-line comment.

\begin{table}[h!]
	\centering
	\begin{tabular}{lrr}
		\toprule
		\textbf{Type of}                        & \textbf{\#Instances}			& \textbf{\#Instances}\\
		\textbf{instance}						& \textbf{Pre-training}			& \textbf{Fine-tuning}\\
		\midrule
		Inner comment(s) only   \textbf{(D1)}               & 45,764       & 33,590 \\
		Javadoc comment only       \textbf{(D2)}              &  232,121    &  115,904      \\
		Javadoc \& Inner comments     \textbf{(D3)}        &      53,667     &      16,282     \\
		\midrule                                     
		\textbf{Total} & 331,552   &    165,776     \\
		\bottomrule
	\end{tabular}
	\caption{Instances used for pre-training and fine-tuning.}
	\label{tab:dataset}
\end{table}

Finally, we removed duplicates, obtaining the study dataset composed of \totalmethods instances, with each instance being a method with its related Javadoc and/or inner comments. 

We randomly split this dataset using $2/3$ of it for pre-training and $1/3$ for fine-tuning. \tabref{tab:dataset} shows the number of instances in each of the two datasets, distinguishing between instances only containing inner comments (\textbf{D1}), only containing Javadoc (\textbf{D2}), and featuring both (\textbf{D3}).

\subsection{Pre-training of T5}\label{subsec:training_strategy}

In the \textit{pre-training} phase, we use a self-supervised task similar to the one used by Raffel \etal \cite{raffel2019exploring}, consisting of masking tokens in natural language sentences and asking the model to guess the masked tokens. Since we want the model to learn how to generate comments given a certain context, we randomly mask 15\% of tokens appearing in the comment-related part of each instance (Javadoc or inner comments). Tokens representing the method code were not masked. The pre-training has been performed for 200k steps.

We also created a new \emph{SentencePiece} \cite{DBLP:journals/corr/abs-1808-06226} (\ie a tokenizer for neural text processing) model by training it on the entire pre-training dataset, in such a way that it can handle both code and comments. We set its length to 32k wordpieces.

\subsection{Fine-tuning of T5}\label{subsec:finetuning_strategy}
Once pre-trained, we fine-tune the T5 model in a multi-task setting, in which the two tasks are represented in our case by the automatic completion of (i) Javadoc comments and (ii) inner comments. Having a single model fine-tuned on these two strongly related tasks could result in an effective transfer learning, in which knowledge gained by the model while learning a task (\eg \JTask) can be transferred to other similar tasks (\eg \ICTask).

\subsubsection{Preparing the Dataset for the Model Fine-Tuning}\label{dataset_finetuning_preparation}
We further process the 165,776 instances selected for the fine-tuning (see \tabref{tab:dataset}) through the following steps.

\textbf{Processing Javadoc instances (datasets D2 and D3 in \tabref{tab:dataset})}.
For the \JTask we assume that the \emph{context} documented by a \JDComment is represented by the whole method. 

Thus, given an instance composed by $\{$\texttt{\JDComment}, \texttt{context}$\}$ (\ie a Javadoc comment followed by the method it documents) we simply swap the position of the two elements and add a special separation token $<$\texttt{sep}$>$ to delimit the comment, obtaining an instance in the form: $\{$\texttt{context}$<$\texttt{sep}$>$\texttt{\JDComment}$<$\texttt{sep}$>$$\}$. The rationale behind this transformation is to ``force'' the model to process the context before predicting the missing parts of the comment. 

Once this is done, we use the method \texttt{sent$\_$tokenize} from the \emph{nltk} library \cite{nltk} to split the \JDComment in each instance into the $y$ sentences composing it. Then, we take the first sentence $s_1$ and remove the remaining $y-1$. Assuming $s_1$ is composed by $n$ tokens, we randomly extract five different integers between 1 and $n-1$, and use them to create five variants of $s_1$ each one having the last $n-m_i$ tokens masked, where $m_i$ is one of the five random integers. 

By training the model on these five masked sentences, the learning is focused on guessing how to finalize an incomplete sentence in a comment the developer is writing. Let us justify and explain this process. First, we remove the $y-1$ following sentences because we assume that a developer writes the comment linearly, starting from the first to the last sentence. Thus, when the developer is writing the first sentence, the remaining $y-1$ do not exist yet. Second, at most $n-1$ tokens can be masked in a sentence composed by $n$ tokens, since at least the first token must be provided by the developer (otherwise, the task would be comment generation rather than completion). Third, the choice of creating five different variants for a given sentence is a tradeoff between experimenting with a different number of masked tokens for each sentence and considering all possible combinations of masked tokens, that would lead to an excessive number of training instances. 

\begin{figure}
	\includegraphics[width=\linewidth]{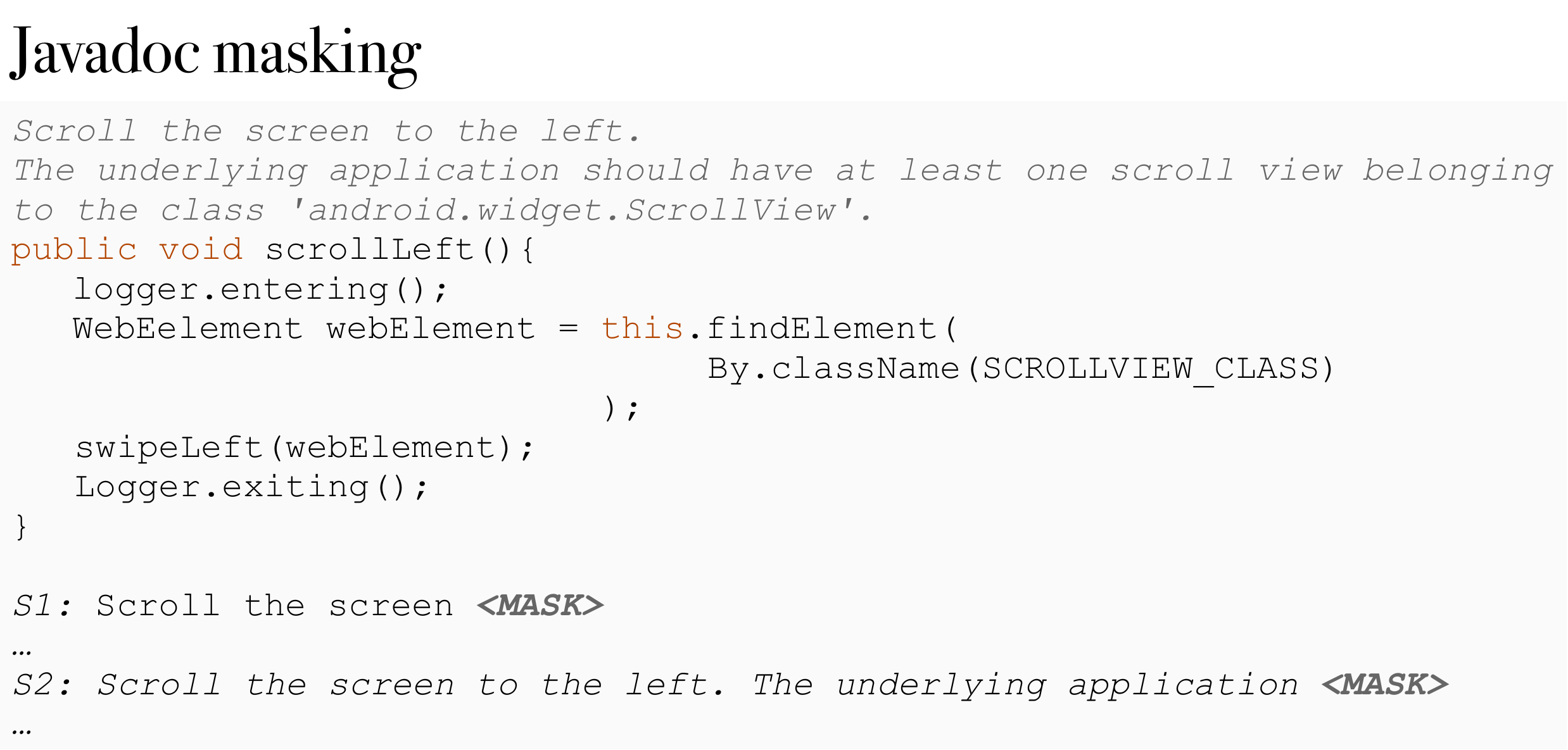}
	\caption{Example of the Javadoc masking process}
	\label{fig:masking}
\end{figure}

Such a process is repeated for all $y$ sentences composing the \JDComment, hiding the sentences following the one under process while keeping the ones preceding it. Thus, for each instance in our fine-tuning dataset, we create up to $y$*5 instances (\ie $y$ sentences with five different sets of masked tokens). Less than five instances are created if \JDComment has less than six tokens, since it would not be possible to mask five different sets of tokens (excluding the first one). \figref{fig:masking} shows an example of masking performed on a single instance. The original instance is reported on top of the figure, and two sentences compose its \JDComment. This results in the creation of 10 fine-tuning instances. Due to space constraints, \figref{fig:masking} only shows one of the instances generated for each sentence.

\begin{figure}
	\includegraphics[width=\linewidth]{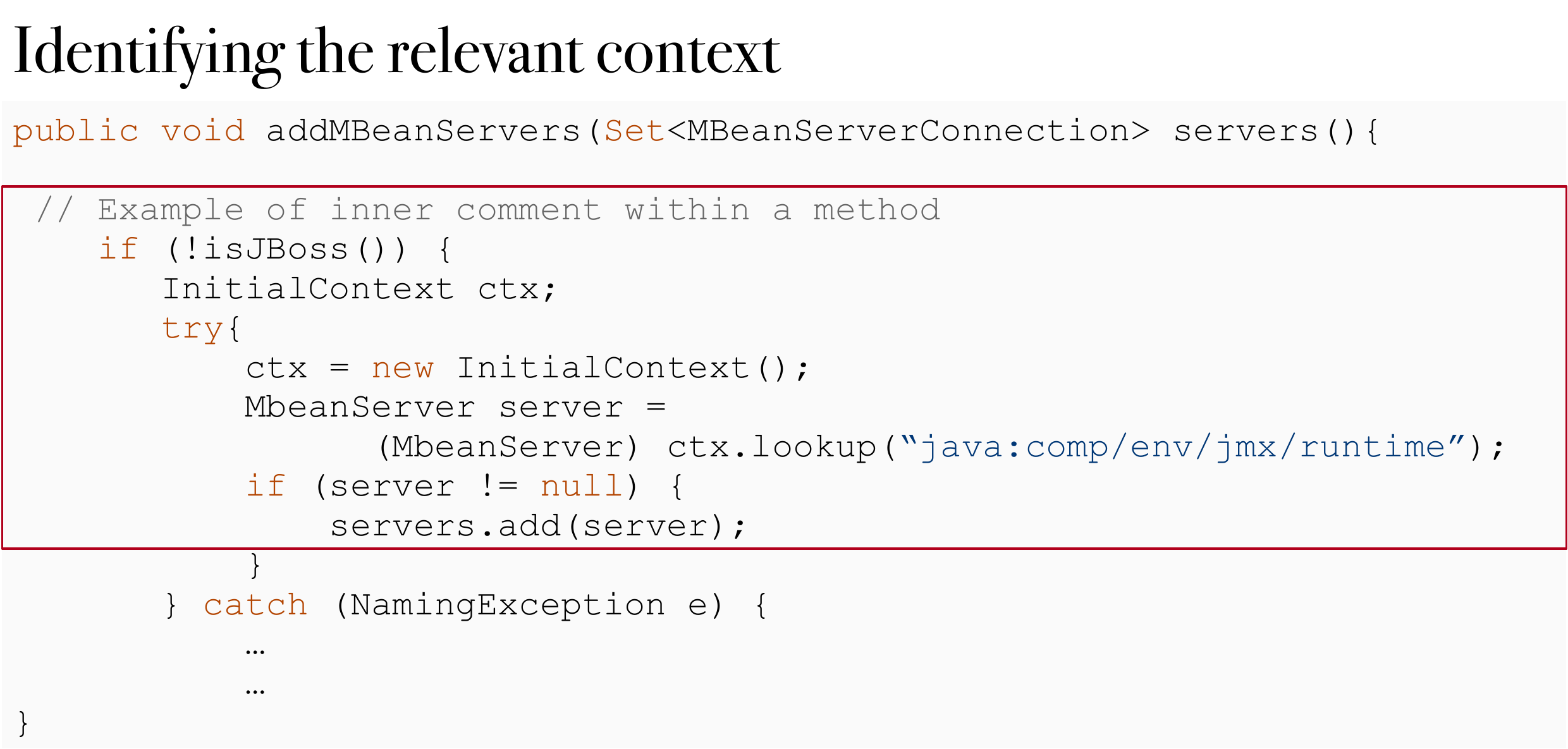}
	\caption{Identification of relevant context of an inner comment}
	\label{fig:inner}
\end{figure}

\textbf{Processing Inner-comment instances (datasets D1 and D3 in \tabref{tab:dataset})}. 
For the \ICTask the first step to perform when preparing the fine-tuning dataset is the identification of the \emph{context} relevant for a given inner method. We define the following heuristic to identify, given an \IComment in a specific instance, the context that can help the model in understanding how to support \IComment completion. We use \figref{fig:inner} as a running example to explain the heuristic, showing an example of instance having a single \IComment. Starting from the line $l_\IComment$ containing it, we expand the context both above and below it until specific conditions are met. 

In particular, while expanding above and below $l_\IComment$, we stop when we find one of the following: (i) an empty line, (ii) a closing curly brace, or (iii) another code comment. In both cases, we do not expand the context out of the method (\eg if none of the above conditions is met while expanding above $l_\IComment$, we stop at the method signature). In the example shown in \figref{fig:inner}, there is no above context since we immediately hit an empty line, while the context below is stopped when we find the first closed curly brace. It is important to clarify that the \emph{context} we identify is not necessarily the part of code documented by \IComment. However, our interest is to provide the model with the relevant code surrounding \IComment, to allow it exploiting useful information to support the comment completion. Once linked each \IComment to its context, we perform the same processing previously described for the \JDComment instances (\ie splitting the comment into sentences and creating five variants of each sentence, each having a different number of tokens masked at the end of it). 

\begin{table}[h!]
	\centering
	\begin{tabular}{lrrr}
		\toprule
		\textbf{Data sources}                        & \textbf{Train}  & \textbf{Eval} & \textbf{Test}\\
		\midrule
		\JTask                  							  & 1,398,135         & 174,624   &  175,084\\
		\ICTask                     					     &  272,944           & 34,705     &  34,138 \\
		\hline	
		\textbf{Total}								       & 1,671,079         & 209,329 & 209,222\\

		\bottomrule
	\end{tabular}
	\caption{Instances used for the fine-tuning}
	\label{tab:finetune_dataset}
\end{table}

\subsubsection{Dataset Splitting} \label{sub:dataset}
\tabref{tab:finetune_dataset} shows the fine-tuning dataset we obtained from the above-described process. We split it into 80\%-10\%-10\% for train, test and validation, respectively. The dataset for the \JTask dominates, in terms of size, the one for the \ICTask. 

This could result in an unbalanced effectiveness of the model for the two tasks. In other words, the model could perform better on the \JTask and being less effective on \ICTask. However, as pointed out by Arivazhagan \etal \cite{DBLP:journals/corr/abs-1907-05019}, there is no free lunch in choosing the balancing strategy when training a multi-task model, with each strategy having its pros and cons (\eg oversampling of less represented datasets negatively impacts the performance of the most representative task). For this reason, we decided not to perform any particular adaptation of our training set but to follow the true data distribution when creating batches.

\subsection{Decoding Strategy}
The given output layer's values allow different possible decoding strategies to generate the output token streams. We adopt a \textit{greedy decoding} strategy since it we believe it is better suited for the problem we are tackling. Indeed, it provides the developer with the most likely completion (rather than with a set of completions as, for example, a \textit{beam search} would do. Indeed, with multiple options a developer would likely spend more time selecting among different tool suggestions rather than writing the comment themself. 

\subsection{Hyperparameter Tuning}
We rely on the configurations used by Mastropaolo \etal \cite{Matropaolo:icse2021}. Concerning the pre-training, we do not tune the hyperparameters of the T5 model because the pre-training step is task-agnostic, and this would provide limited benefits. Instead, we experiment with four different learning rates schedule for the fine-tuning phase, using the configurations reported in \tabref{tab:hyperparameter:types}.

\begin{table}[h]
	\centering
	\vspace{3mm}
	\begin{tabular}{ll}
		\toprule
		\textbf{Learning Rate Type}   & \textbf{Parameters}\\
		\midrule
		Constant (C-LR)            & $\mathit{LR} = 0.001$        \\
		Slanted Triangular (ST-LR)  
		& $\mathit{LR}_{\mathit{starting}} = 0.001$ \\
		& $\mathit{LR_{\mathit{max}}} = 0.01$\\
		& $\mathit{Ratio} = 32$\\
		& $\mathit{Cut} = 0.1$\\
		Inverse Square Root (ISQ-LR) 
		& $\mathit{LR}_{\mathit{starting}} = 0.01$ \\
		& $\mathit{Warmup} = 10,000$ \\
		Polynomial Decay (PD-LR)    
		& $\mathit{LR}_{\mathit{starting}} = 0.1$\\
		& $\mathit{LR}_{\mathit{end}} = 0.01$\\
		& $\mathit{Power} = 0.5$\\
		\bottomrule
	\end{tabular}
	\caption{Learning-rates tested for hyperparameter tuning}
	\label{tab:hyperparameter:types}
	\vspace{-0.2cm}
\end{table}

We fine-tune the model for 100k steps for each configuration; then, we compute the percentage of perfect predictions (\ie cases in which the model can correctly predict all masked tokens in the comment) achieved on both tasks in the evaluation set. The achieved results reported in \tabref{tab:hyperparameter:results} showed a slight superiority of the \emph{slanted triangular} (column 2) that we use in our study.

\begin{table}[h!]
	\centering
	\vspace{3mm}
	\begin{tabular}{lrrrr}
		\toprule
		\textbf{Task}                  & \textbf{C-LR}    & \textbf{ST-LR}     & \textbf{ISQ-LR}   & \textbf{PD-LR} \\
		\midrule
		\JTask                            &  30.77\%                & 33.60\%    & 31.73\%                    & 30.65\%         \\
		\ICTask                        &  9.38\%                         & 10.55\%     &  9.70\%                      &  9.51\%         \\		
		\bottomrule
	\end{tabular}
	\caption{Hyperparameter tuning results}
	\label{tab:hyperparameter:results}
	\vspace{-0.4cm}
\end{table}
\section{Study Design} \label{sec:design}

\newcommand{\Shared}[1]{$\mathit{Shared}_{\mathit{#1}}$}
\newcommand{\OnlyOurs}[1]{$\mathit{OnlyT5}_{\mathit{#1}}$}
\newcommand{\OnlyBaseline}[1]{$\mathit{OnlyNG}_{\mathit{#1}}$}
\newcommand{\RQ}[1]{RQ\textsubscript{#1}}

\newcommand\rqOne{To what extent can T5 and $n$-gram models be leveraged to support the auto-completion of code comments?}

\newlength\Linewidth
\def\findlength{
	\setlength\Linewidth\linewidth
	\addtolength\Linewidth{-4\fboxrule}
	\addtolength\Linewidth{-3\fboxsep}
}

\newenvironment{researchquestion}
	{\par\begingroup
	\setlength{\fboxsep}{5pt}\findlength
	\setbox0=\vbox\bgroup\noindent
	\hsize=0.95\linewidth
	\begin{minipage}{0.95\linewidth}\normalsize}
	{\end{minipage}\egroup
	   \vspace{5pt}
	\textcolor{gray}{\fboxsep1.5pt\fbox{\fboxsep5pt\colorbox{white}{\normalcolor\box0}}}
	\endgroup\par\noindent
	\normalcolor\ignorespacesafterend}

The \textit{goal} of this study is to experiment the extent to which a T5 model and an $n$-gram model can help developers in writing comments faster by supporting code comment completion. 

In particular, we answer the following research question: \emph{\rqOne}

We answer this research question by using the 209,222 test set instances in \tabref{tab:finetune_dataset} as a context for our study. This means that the two trained models (\ie T5 and $n$-gram model) are run on the same test set instances to predict the masked parts of code comments. The T5 model has been trained as described in \secref{sec:approach}, while in the following we explain how we implemented and trained the $n$-gram models. The code and data used in our study are publicly available \cite{replication}.

\subsection{$N$-Gram Model}
An \textit{n}-gram model can predict a single token following the $n-1$ tokens preceding it. We publicly release the implementation of our $n$-gram model in Python \cite{replication}. We trained the \textit{n}-gram model by using the same instances used to fine-tune the T5 without, however, the masked tokens. We experimented with three different values for $n$ (\ie $n=3$, $n=5$ and $n=7$). Even though the $n$-gram model is meant to predict a single token given the $n-1$ preceding tokens, we designed a fair comparison for the scenario in which we mask more than one token. In particular, we use the $n$-gram model in the following way: Let us assume that we are predicting, using a $3$-gram model, how to complete a sentence having five tokens \texttt{T}, of which the last two are masked (\texttt{M}): $<$\texttt{T$_1$}, \texttt{T$_2$}, \texttt{T$_3$}, \texttt{M$_4$}, \texttt{M$_5$}$>$. We provide as input to the model \texttt{T$_2$} and \texttt{T$_3$} to predict \texttt{M$_4$}, obtaining the model prediction \texttt{P$_4$}. Then, we use \texttt{T$_3$} and \texttt{P$_4$} to predict \texttt{M$_5$}, thus obtaining the predicted sentence $<$\texttt{T$_1$}, \texttt{T$_2$}, \texttt{T$_3$}, \texttt{P$_4$}, \texttt{P$_5$}$>$. Basically, all predictions are joined to predict multiple contiguous tokens.

We experimented with the different values of $n$ by running the models on the evaluation set, taking the best one (\ie \textit{5}-gram) and comparing its performance with that of the T5 model. We report the results achieved for the \textit{3}-gram and \textit{7}-gram models in our replication package \cite{replication}.

\subsection{Evaluation Metrics and Data Analysis} \label{sub:metrics}
We compare the T5 and $5$-gram models using six metrics.

\textbf{Perfect predictions}: This metric measures the percentage of cases (\ie instances in the test set) in which the sequence predicted by the model equals the oracle sequence. Since we want to investigate the extent to which the experimented techniques can actually support comment completion, we compute the perfect predictions when the technique is only required to guess the first masked token, the first two, the first three, \etc For example, if we assume that in the previous prediction $<$\texttt{T$_1$}, \texttt{T$_2$}, \texttt{T$_3$}, \texttt{P$_4$}, \texttt{P$_5$}$>$ \texttt{P$_4$} is correct while \texttt{P$_5$} is wrong, we will consider this as a perfect prediction only when looking at the first token to predict, and as a wrong one when looking at the first two. 

We compute the percentage of perfect predictions when trying to predict the first $k$ masked tokens, with $k$ going from 1 to 10 at steps of 1 (\ie 1, 2, 3, \etc) and for the most challenging scenario in which $k>10$ (\ie more than 10 masked tokens must be correctly predicted to consider this prediction as a perfect one).

\textbf{BLEU score \cite{Papineni:2002}:} This metric measures how similar the candidate (predicted) and reference (oracle) texts are. Given a size $n$, the candidate and reference texts are broken into $n$-grams, and the algorithm determines how many $n$-grams of the candidate text appear in the reference text. The BLEU score ranges between 0 (the sequences are completely different) and 1 (the sequences are identical). For both the tasks, we compute the BLEU-\{1, 2, 3, 4\} and their geometric mean (\ie BLEU-A). Due to the way in which the BLEU-$X$ is computed (\ie at least $X$ tokens must be part of the prediction task) we only compute the BLEU-A metric when the number of tokens for a given prediction is at least 4. As done for perfect predictions, we report results for different values of $k$.

\textbf{Levenshtein distance \cite{levenshtein1966binary}:} To understand the effort needed by developers to convert a prediction generated by the model into a correct comment, we compute the Levenshtein distance at word-level, \ie the minimum number of word edits (insertions, deletions or substitutions) needed to convert the predicted comment into the reference one. Also in this case, we report results for different values of $k$.

\textbf{Overlap metrics:} We also compute the complementarity between the T5 and the $n$-gram model. Let $\mathit{PP}_{\mathit{T5}_t}$ and $\mathit{PP}_{\mathit{NG}_t}$ be the sets of perfect predictions achieved by T5 and the \textit{n}-gram model, where $t$ $\in$ $\{\ICTask, \JTask\}$. 
We compute the following metrics:

\footnotesize
$$
\mathit{Shared}_{\mathit{t}} = \frac{|\mathit{PP}_{\mathit{T5}_t} \cap \mathit{PP}_{\mathit{NG}_t}|}{|\mathit{PP}_{\mathit{T5}_t} \cup \mathit{PP}_{\mathit{NG}_t}|}
$$

\noindent\begin{minipage}{.5\linewidth}
	$$
	\mathit{OnlyT5}_{\mathit{t}} = \frac{|\mathit{PP}_{\mathit{T5}_t} \setminus \mathit{PP}_{\mathit{NG}_t}|}{|\mathit{PP}_{\mathit{T5}_t} \cup \mathit{PP}_{\mathit{NG}_t}|}
	$$
\end{minipage}%
\begin{minipage}{.5\linewidth}
	$$
	\mathit{OnlyNG}_{\mathit{t}} = \frac{|\mathit{PP}_{\mathit{NG}_t} \setminus \mathit{PP}_{\mathit{T5}_t}|}{|\mathit{PP}_{\mathit{T5}_t} \cup \mathit{PP}_{\mathit{NG}_t}|}
	$$
	\smallskip
\end{minipage}
\normalsize

\Shared{t} measures the percentage of perfect predictions shared between the two compared approaches, while \OnlyOurs{t} and \OnlyBaseline{t} measure the percentage of cases in which the perfect prediction is only achieved by T5 or the $n$-gram model, respectively, on the task $t$.

\textbf{Confidence Analysis}: Both models provide, together with the generated prediction, a score between 0 and 1 indicating the confidence of the prediction, with 1 being the maximum confidence. We check whether the \emph{confidence} of the predictions can be used as a reliable proxy for their ``quality''. If this is the case then, a possible implementation of these models into a tool could take advantage of this proxy to automatically filter out low-confidence predictions. For each model, we compute the confidence level for the two sets of ``perfect'' and wrong predictions comparing their average confidence.

\textbf{Qualitative analysis of the predictions:} To better understand the strengths and weaknesses of the models, we analyze more closely the generated predictions. First, we check what type of words on two models are able to correctly predict. 

We do this by performing a \textit{Part-of-Speech (POS) TAG} analysis. For each test set instance we check the POS category of each masked word using 12 POS categories \cite{tagset} (\eg adjective, adverb, noun). Then, for each POS category, we compute for both models the percentage of times they were able to correctly predict it. Such an analysis is useful to understand whether the words correctly predicted by the models are mostly trivial ones (\eg determiners such as ``the'', ``a'') or also more challenging words representing, for example, nouns. On top of that, we discuss examples of predictions made by the two models.


A statistical comparison between the T5 and the $n$-gram model is performed using the McNemar's test \cite{mcnemar} and Odds Ratios (ORs) on the perfect predictions they can generate.


\begin{figure*}
	\centering
	\includegraphics[width=0.8\linewidth]{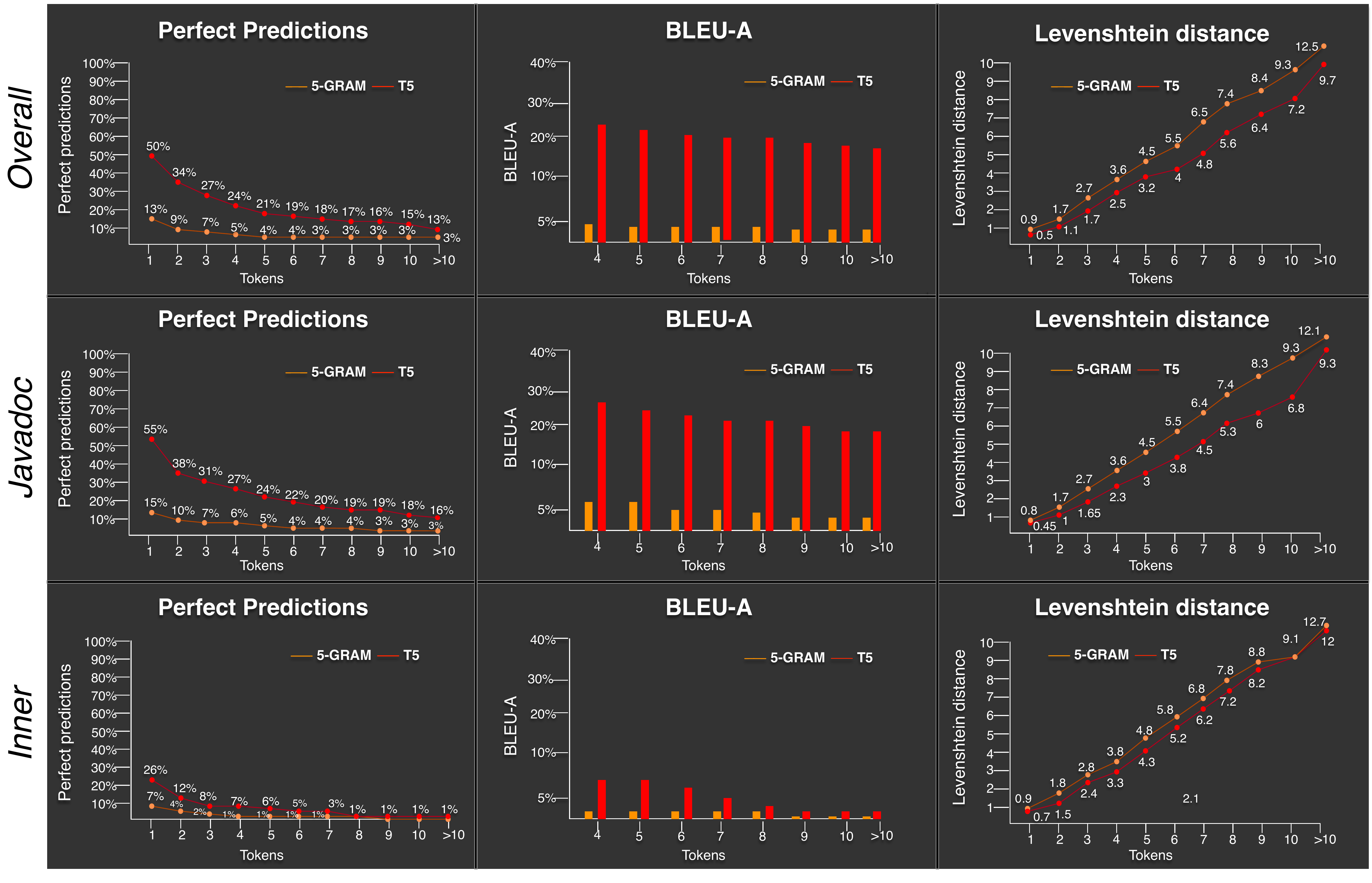}
	\vspace{-0.15cm}
	\caption{Performance of the T5 model against the $5$-gram model}
	\label{fig:results}
	\vspace{-0.3cm}
\end{figure*}

\section{Results Discussion} \label{sec:results}
\figref{fig:results} depicts the results achieved by the T5 and $5$-gram model in terms of perfect predictions, BLEU-A score, and Levenshtein distance computed for predictions of different lengths ($k$). The results for the other metrics (\eg BLEU-1 to BLEU-4) can be found in our replication package \cite{replication}. The middle and bottom parts of \figref{fig:results} show the results achieved in the \JTask and \ICTask, respectively, while the top part aggregates the results of the two datasets. 

In both the evaluated tasks (\ie \JTask and \ICTask) as well as overall, T5 outperforms the $5$-gram model by a significant span for all metrics considered in our study. This is especially true when the two models are required to predict a limited number of tokens ($k\leq6$) following the ones already written by the developer in the code comment. For example, when only the subsequent word must be predicted (\ie $k=1$), T5 can achieve more than 50\% of perfect predictions in the \JTask and more than 25\% in the \ICTask. The $5$-gram model, in both scenarios, achieves less than 16\% of perfect predictions. 

The difference in performance is particularly remarkable for the \JTask, in which the T5 model achieves better results compared to the \ICTask. Such a finding might be due to two important factors. First, as explained in \secref{sub:dataset}, the fine-tuning dataset used for the \JTask is larger than the one used for the \ICTask, thus likely providing more knowledge to the model about the vocabulary usually adopted by developers in Javadoc comments and, more in general, about this specific task. Second, the Javadoc has, by its nature, a more regular structure making use of tags (\eg \texttt{@param}) that could help the model in better predicting the comment, especially given the fact that the T5 model exploits the relevant code context during the prediction. Still, even on the \ICTask the T5 model achieves three times more perfect predictions of the $5$-gram when predicting up to seven tokens (bottom-left corner of \figref{fig:results}).

When the number of tokens to predict increases, the gap in performance between the two approaches gets thinner. In the most complex scenario in which the models predict more than 10 tokens in the comment, the T5 achieves, overall, 13\% of perfect predictions against the 3\% of the $5$-gram model. 

The difference is smaller for the \ICTask, with the best approach (T5) achieving only 1\% of perfect predictions. 

The McNemar's test always indicates significant differences in terms of perfect predictions ensured by the T5 and the $5$-gram model, with ORs ranging between 8.04 for the \ICTask and 17.56 for the \JTask (OR=16.79 for the overall dataset). The better performance of T5 is confirmed by the other evaluation metrics we adopted, namely the BELU-A and the Levenshtein distance. The BLEU-A gap is up to five times in favor of the T5 over the $5$-gram, confirming a substantial difference in performance between the two models. On the \JTask, T5 always achieves a BLEU-A score higher than at least $\sim$16\% as compared to that achieved by the $5$-gram, independently of the number of tokens the two models are asked to predict. As already observed for the perfect predictions, the BLEU-A gap is much smaller in the \ICTask, however still showing a plus $\sim$5\% in favor of T5 up to six tokens. The difference in performance tends to decrease while increasing the number of tokens to predict.

By focusing on the Levenshtein distance (right part of \figref{fig:results}, the lower the better), we can observe that, as expected, the number of token-level edits needed to convert a prediction into the reference one tends to increase for both models when more tokens are predicted. An analysis of both tasks (\ie \JTask and \ICTask) points out that T5 requires a developer intervention in a lower number of cases than $5$-gram. However, a clear conclusion can be drawn by looking at the three graphs in the right part of \figref{fig:results}: When the two models are not able to generate a perfect prediction, the effort required by developers to convert the prediction into the comment they actually want to write might be too high. For example, when predicting the next five tokens the developer is likely to type, the T5 requires, on average, to changes to 3.2 of the predicted tokens.

\begin{table}
	\centering
	\begin{tabular}{l r r r}
		\toprule
		\emph{Task} ($d$)      & \Shared{t}     & \OnlyOurs{t}    & \OnlyBaseline{t} \\
		\midrule
		\JTask                  & 17.06\%        & 75.87\%            & 7.07\%          \\
		\ICTask                 & 19.70\%        & 67.78\%         & 12.52\%          \\
		\bottomrule
	\end{tabular}
	\caption{Perfect predictions overlap between T5 and \textit{5}-gram} 
	
		\label{tab:overlap}
\end{table}

An important point to discuss in the comparison between T5 and $5$-gram is the different datasets used for their training. Indeed, T5 benefited of a pre-training phase in which it exploited additional code that was not made available to the $5$-gram  during training. Thus, we performed an ablation study on the T5 model by removing its pre-training step and checking to what extent its superior performances are due to the performed pre-training. While details can be found in our replication package \cite{replication}, we can summarize our findings as follows: The pre-training phase increases the performance of the T5 in terms of perfect predictions in a range going from $\sim$0.5\% to $\sim$2\%, depending on the task and on the $k$ value. Thus, while pre-training is beneficial, the performance of the T5 are still better than those of the $5$-gram model even when both models are only trained on the fine-tuning dataset.  

\tabref{tab:overlap} reports the results of the overlap metrics we computed (see \secref{sub:metrics}). For the \JTask, only 17.06\% of the perfect (\ie correct) predictions are shared among the two models, while 75.87\% are correctly generated only by the T5 model. The \textit{5}-gram is responsible for the remaining 7.07\% of perfect predictions, that are missed by the T5. This shows, at least for the \JTask, a limited (but existing) complementarity between the models. Similar results are achieved for the \ICTask. The two models share 19.70\% of perfect predictions, with 67.78\% of them correctly predicted by T5 only. The \textit{5}-gram model contributes the remaining 12.52\% again showing some complementarity between the models.

\begin{figure*}
\centering
	\includegraphics[width=0.8\linewidth]{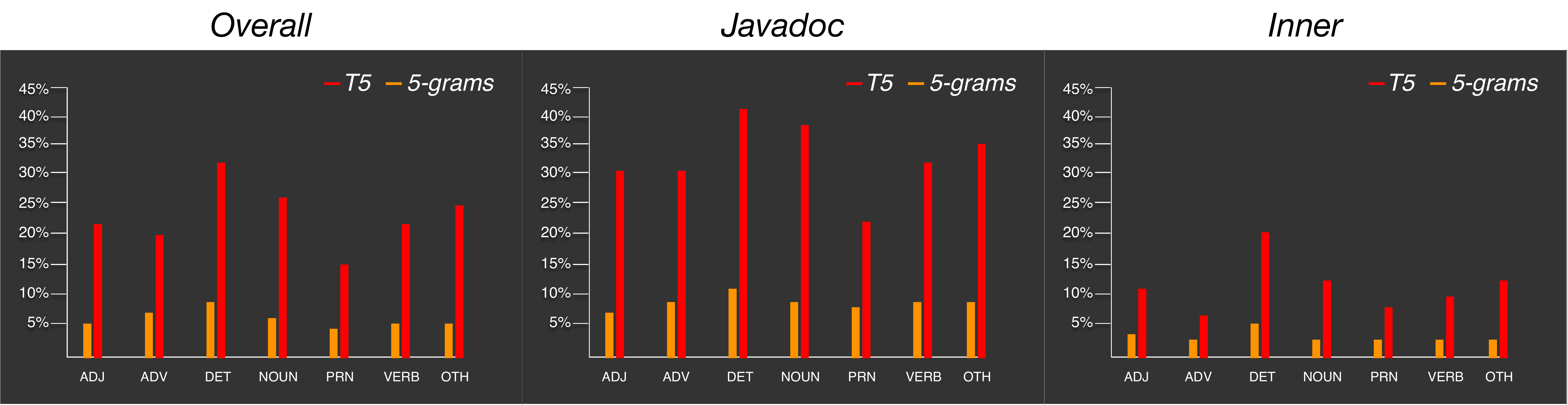}
	\caption{Part-of-speech tag analysis: ADJ=adjective, ADV=adverb, DET=determiner, PRN=pronoun, OTH=others which includes \textit{conjunction, number, particle, adposition, and X}, where X are words that cannot be assigned a part-of-speech category}
	\label{fig:pos_analysis}
	\vspace{-0.4cm}
\end{figure*}

\begin{figure}
	\centering
	\includegraphics[width=0.8\linewidth]{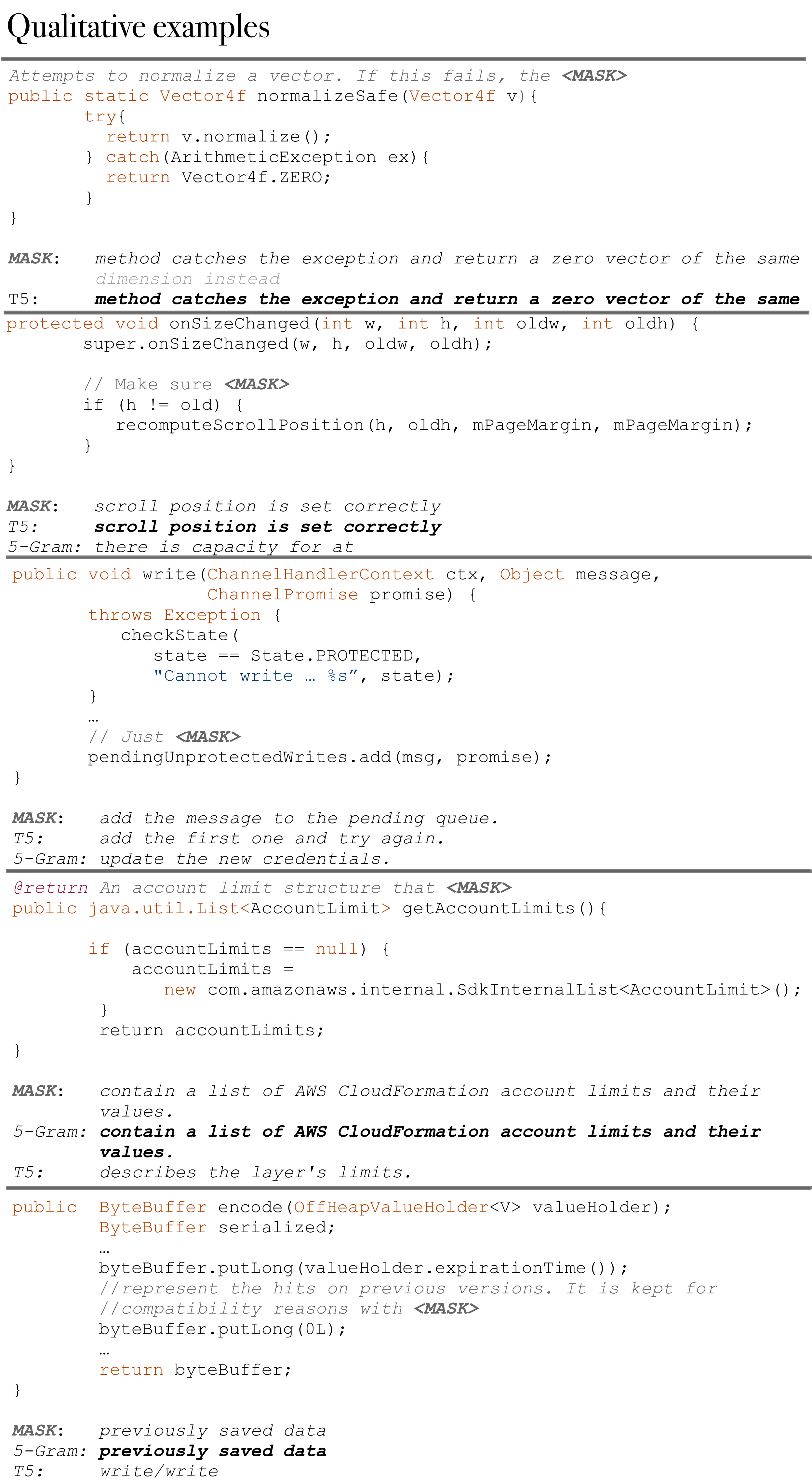}
	\caption{Qualitative examples.}
	\vspace{-0.3cm}
	\label{fig:qualitative}
\end{figure}

\figref{fig:pos_analysis} shows the POS analysis (\ie percentage of correct predictions of each POS type) that confirms the superior performance of T5 across all investigated POS categories: Independently from the type of word to predict, the T5 outperforms the \textit{5}-gram model. Also, the performances on the \JTask are, as expected, superior. Not surprisingly, 
determiners are the ones having the highest percentage of correct predictions. Nevertheless, POS types like nouns, adjective, and verbs which are certainly more challenging to predict still exhibit a good percentage of correct predictions. This analysis, combined with the previous one showing the performance of the model at different values of $k$, shows that the perfect predictions obtained by the two models (and in particular by the T5) are not only the result of trivial single word ($k=1$) predictions involving simple POS types, but also include more challenging prediction scenarios. \figref{fig:qualitative} reports five qualitative examples of predictions performed by both models: the first two are successful predictions made \label{key}only by the T5 for the \JTask and the \ICTask, respectively.

Note that for the first one, the $5$-gram does not generate any prediction likely due to the fact that the $4$-gram provided as input was never encountered in the training set. The T5, instead, correctly guesses the subsequent 12 tokens in the comment out of the 14 we masked (thus, this is a perfect prediction when considering $k=12$). The third qualitative example is a prediction failed by both techniques, with the comment generated by the T5 model being more similar to the reference one. Finally, the two predictions at the bottom represent cases in which the $5$-gram model correctly completes the comment while the T5 comment fails. Also in this case the first of the two examples is taken from the \JTask, while the second represents an \ICTask. The complete list of predictions is publicly available \cite{replication}.

Overall, our quantitative analysis showed the superiority of the T5 model in the code comment completion task. However, it is important to mention that the T5 model exploits during the prediction the \emph{context} we provide as input (\ie the code likely to be relevant for the specific comment). This means that, from a practical point of view, such a model can be exploited for code comment completion only assuming that the developer first writes the code and, then, the comment to document it. Clearly, this is not always the case and limits the applicability of the T5 model we experimented with. Such a problem is instead not present in the $n$-gram model, that can always be applied as long as $n-1$ tokens are present before triggering the prediction of the $n^{th}$ token. A tool integrated into an IDE to support code comment completion could exploit both models: The $n$-gram model could be triggered if no context can be captured for the comment being written, while the T5 can perform the prediction when relevant code  is present.

\begin{figure*}
	\centering
	\includegraphics[width=0.8\linewidth]{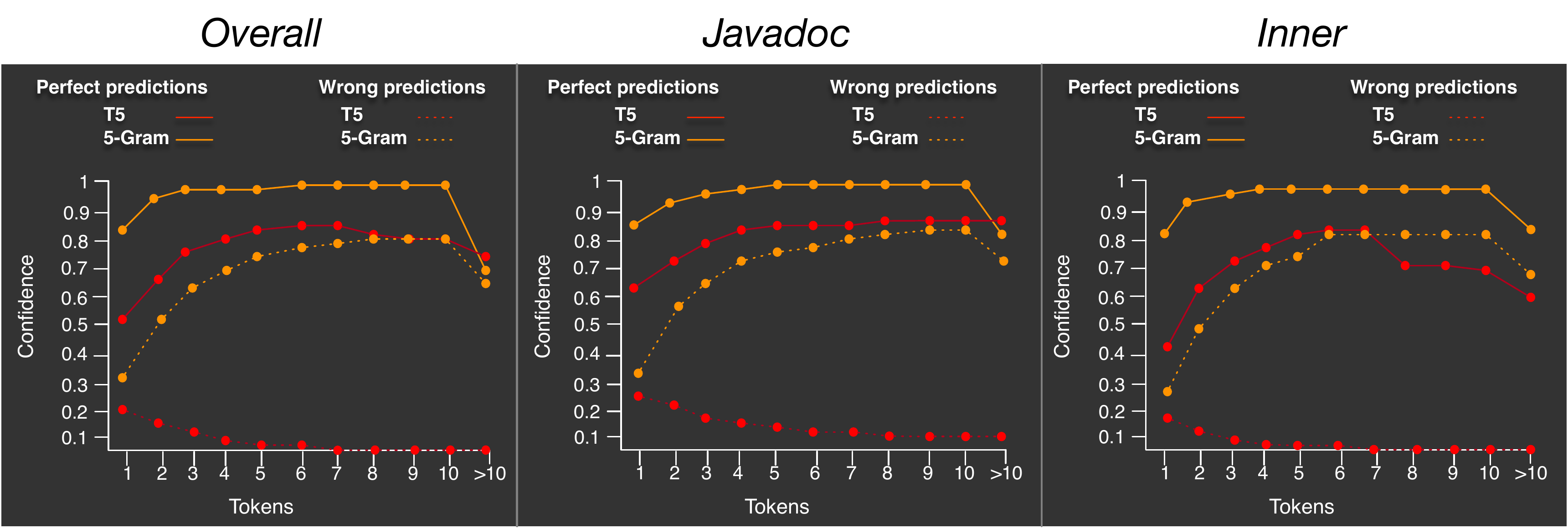}
	\caption{Confidence level in relation to the length of the predicted tokens}
	\vspace{-0.3cm}
	\label{fig:confidence_analysis}
\end{figure*}

\figref{fig:confidence_analysis} depicts the average \emph{confidence} level for perfect (continues lines) and wrong (dashed) predictions made by the T5 (red lines) and the \textit{5}-gram (orange) model. As it can be seen, the confidence of both models is a good proxy for the quality of the predictions. This is particularly true for the T5, for which we can see as the correct predictions tend to have a confidence greater than 0.5 for all $k$ values, while the wrong predictions have confidence approaching 0.0 when $k$ increases. Thus, a threshold based on confidence could be used in IDE tools to avoid recommending predictions likely to be wrong. 

Finally, it is worth mentioning that, as compared to the task of code comment generation (\ie generating a comment from scratch given a code as input), the performance achieved in terms of perfect prediction is substantially higher. Indeed, when looking at a similar model (T5) experimented in the context of Javadoc generation, it achieved $\sim$10\% perfect predictions \cite{Matropaolo:icse2021}. The results in \figref{fig:results} show that, depending on the length of the prediction on the \JTask, for code comment completion, the same model can achieve $\sim$16-55\% perfect predictions. Thus, it might be more feasible for such a (simpler) problem to develop tools that can already support developers in their everyday coding activities.


\section{Threats to Validity} \label{sec:threats}

\textbf{Construct validity.} While building our dataset, we made three important assumptions. First, by providing the code context as input to the T5, we assume that the comment is written by developers when the code is already implemented, which is not always the case. Still, this does not invalidate the reported results, but it means, from a practical point of view, that comment completion recommendations could be triggered by the T5 only when comments are added after the code is written. Second, when fine-tuning the T5 model, we hid from the comment under analysis the sentences following the ones in which we masked tokens. Thus, we are assuming that the comment is written linearly (\ie one sentence after the other) which, again, is not always the case. Third, in our study we use the original comment written by developers as oracle, assuming that it represents a valuable reference for the experimented model. Also this assumption, while done in many previous works \cite{Hu:icpc2018,iyer:acl,Allamanis:2016,Hu:emse2020,haque:2020}, might be wrong.

\textbf{Internal validity.} An important factor that influences DL performance is hyperparameters tuning. For the pre-training phase, we used the default T5 parameters selected in the original paper \cite{raffel2019exploring}. For the fine-tuning, we did not change the model architecture (\eg number of layers), but we experimented with different learning rates. We are aware that a more extensive calibration would likely produce better results.

\textbf{External validity.} Our study involved $\sim$500k Java methods, thus ensuring a good generalizability at least for Java code. Our results cannot be  generalized to other languages. 
\section{Related Work} \label{sec:related}

Our work is related to approaches (semi)automatically documenting source code. They can be roughly classified into approaches relying on summarization techniques (\eg \cite{Haiduc:WCRE10,Rodeghero:ICSE14}), mining crowd knowledge (\eg \cite{Rahman:SCAM15,Vassallo:ICPC14}) and using supervised learning techniques (\eg \cite{LeClair:icse2019,haque:2020}). We briefly discuss the first two categories, while we focus more on the latter being it closer to our research.

\subsection{Code Summarization}
These techniques can generate code summaries in the form of (i) bag of words representing the main responsibilities of the code \cite{eddy2013evaluating, Haiduc:SPC2010, Rodeghero:ICSE14, abebe2015extraction}; (ii) distilled code, generated by hiding lines not considered fundamental for the code comprehension \cite{Fowkes:RALS14}; and (iii) natural language text, trying to describe the code functionality as humans would do \cite{Sridhara:ASE10, Rastkar:ICSM11}. The first two categories are considered \emph{extractive} approaches, since they synthesize the summary by extracting the most important elements from the original input, while the latter is representative of \emph{abstractive} approaches, that can include in the summary information not present in the code to document. 

Both extractive and abstractive techniques have been used to document code components at different granularity levels, such as method (\eg \cite{Haiduc:WCRE10, Rodeghero:ICSE14, Sridhara:ASE10, McBurney:ICPC14-2, Dragan:ICSM06}), 
method parameters (\eg \cite{sridhara2011generating}),
method usages (\eg \cite{McBurney:ICPC14, McBurney:TSE16})
class (\eg \cite{Fowkes:RALS14, Dragan:ICSM10, Moreno:icpc2013}), unit tests \cite{Li:ICPC18}, and code snippets (\eg \cite{sridhara2011automatically, sridhara2012automatic}).

The abstractive approaches working at method/code snippet level are the most relevant for our research, since we aim at auto completing code comments that are (i) related to a single method or to a part of it; and (ii) written in natural language. Despite these similarities the tackled problem is different.

\subsection{Mining Crowd Documentation}
Another category of approaches relies on mining and processing online resources to aid program comprehension. These approaches have been used to recommend API usage examples \cite{li2018learning, Holmes:ICSE05, Stylos:VLHC06, Murphy:TSE06, Moreno:ICSE15, Subramanian:ICSE14,Jiang:ICPC17, Treude:ICSE16, Robillard:EMSE15, Buse:ICSE12, Bajracharya:FSE10, Mcmillan:TOSEM13, Ponzanelli:MSR14, Ponzanelli:ICSE17, Xie:msr06, zhong2009mapo, Wang:msr13}, fragments of textual/video tutorials relevant for a coding task at hand \cite{petrosyan2015discovering, ponzanelli2016too, ponzanelli2016codetube}, discussions on developers' communications, Q\&A websites/forums relevant for a given code \cite{Treude:ICSE16, Ponzanelli:WCRE13, Ponzanelli:MSR14}, and relevant pieces of information in API reference documentation \cite{Robillard:EMSE15}.

Closer to our work are the studies leveraging crowd knowledge (\eg \cite{Vassallo:ICPC14, Rahman:SCAM15}), or pre-existing documentation (\eg \cite{wong2015clocom, aghaj:2019a}) to document source code. For instance, Rahman \etal \cite{Rahman:SCAM15} implemented \emph{CodeInsight} a tool to mine insightful comments from \emph{StackOverflow} discussing bugs or improvement tips for a code snippet at hand. Wong \etal \cite{wong2015clocom, wong2013autocomment} leveraged existing code comments (within a set of input projects) to automatically generate comments for a target project with an approach called \emph{ColCom}. In a nutshell, their approach reuses code comments extracted from input projects for type-1 and type-2 clones found in the target project. More recently, Aghajani \etal \cite{aghaj:2019a} presented \emph{ADANA}, an approach to automatically inject code comments describing a given piece of Android code at the granularity level intended by the developer. \emph{ADANA} reuses the descriptions of semantically similar code snippets retrieved and processed from \emph{GitHub Gist} and \emph{StackOverflow}.

These techniques could be used to complement and refine the recommendations generated by the models we experimented with. For example, once identified code comments in which the T5 model fails to suggest correct completions, crowdsourced approaches could be used to retrieve similar comments wrote by developers in a similar context (\ie similar code to document) to strengthen the T5 training set on that specific scenario.

\subsection{Machine Learning for Comment Generation \& Completion}
We focus on approaches exploiting machine learning to summarize and document source code, not discussing works done in related but different areas (\eg summarizing code changes \cite{rastkar:icse2013}, suggest descriptive method names \cite{Allamanis:2016}).

Nazar \etal \cite{NAZAR:504} recruited four students with development experience asking them to extract, from a corpus of 127 code snippets, the code lines needed to summarize each snippet. Then, ten developers were asked to inspect the lines selected as relevant for the summary and extract features characterizing them. A total of 21 features have been defined and used to train a SVM classifier able to achieve a precision of 82\% in selecting relevant code lines for a code snippet.

Ying and Robilliard \cite{Ying:FSE13} presented an approach using supervised learning to summarize code examples. 

The supervised algorithm exploits features extracted from the statements composing the code examples. The authors compute the accuracy of their approach by contrasting the generated summaries against those in a manually built oracle, reporting a 71\% precision. The same authors also empirically investigated developers write code summaries \cite{Ying:FSE14}.

DL models are on the rise for the automatic documentation of code. Iyer \etal \cite{iyer:acl} presented \emph{CODE-NN}, a model producing natural language summaries describing C\# code and SQL queries. Zheng \etal \cite{zheng2017code} proposed \emph{Code Attention}, an attention module to translate code to comments. It exploits domain features of code snippets to infer their structure.

Liang \etal \cite{liang2018} used a Recursive Neural Network named Code-RNN to describe the structural information of source code and produce natural language comments. Their Code-RNN takes advantage of a novel GRU cell (Code-GRU) designed for code comments generation. The authors reported the quality of the generated comments by computing the ROUGE score \cite{lin2004rouge} and showing the superiority of the proposed approach as compared to competitive techniques.

Hu \etal \cite{Hu:icpc2018} use a Deep Neural Network (DNN) to automatically generate comments for a given Java method. To train the DNN, the authors mine $\sim$9k Java projects hosted on GitHub by collecting pairs of $\langle$method, comment$\rangle$, where ``comment'' is the first sentence of the Javadoc linked to the method. To assess the effectiveness of their technique, the authors computed the BLEU-4 score \cite{Papineni:2002}, showing the superiority of their approach with respect to the competitive technique presented in \cite{iyer:acl}. 

LeClair \etal \cite{LeClair:icse2019} presented a neural model combining the AST source code structure and words from code to generate coherent summaries of Java methods. The approach, tested on 2.1M methods, showed its superiority as compared to the previous works by Hu \etal \cite{Hu:icpc2018} and Iyer \etal \cite{iyer:acl}.

Haque \etal \cite{haque:2020} presented an approach aimed at documenting Java methods through an encoder-decoder architecture and representing an improvement of the work by LeClair \etal \cite{LeClair:icse2019}. Their model leverages multiple information about the method to document, and in particular: (i) the source code of the method, as a flattened sequence of tokens representing the method, (ii) its AST representation, and (iii) the ``file context'', meaning the code of every other method in the same file. The authors show that adding the contextual information as one of the inputs substantially improves the BLEU score obtained by deep learning techniques. Finally, in a recent work, Mastropaolo \etal \cite{Matropaolo:icse2021} showed that a T5 Model \cite{raffel2019exploring} properly pre-trained and fine-tuned can achieve better performance than the technique presented in \cite{haque:2020}, generating comments ``as humans would do'' in $\sim$10\% of cases.



To summarize, even the most positive results in the literature show major limitations for automatically generating code comments. For this reason, we tackled the ``simpler'' comment completion problem proposed by Ciurumelea \etal \cite{saner2020}. 

The authors mine the top-1000 most starred GitHub python projects in November 2018. Then, they pre-process the extracted data and train three different models. 

The first is a \textit{Sequential Model} taking as input a sequence of Python docstring tokens and trained to predict the single token following that sequence. The second model takes as input a sequence of the Python docstring concatenated with structural information (\ie the method signature). Also in this case, it is trained to predict the next token following the docstring provided as input. Finally, a \textit{Context Model (full body)} is trained for the same task, with the structural information provided to the method represented in this case by the complete method body. The authors show that the third model is the one providing the best results. 

Our study stems from the work by Ciurumelea \etal \cite{saner2020}. However, it is performed in a different context, investigating the actual need for a DL-based model over a simpler $n$-gram model while looking at more challenging prediction tasks.

\section{Conclusions and Future Work} \label{sec:conclusions}

We presented an empirical study comparing two different techniques, namely the T5 and the $n$-gram model, in the task of code comment completion (\ie autocomplete a code comment the developer started writing). The two models are different in nature, with the T5, based on deep learning, exploiting as information to support the completion of a comment $C$ not only the first tokens typed by the developer while writing $C$ but also a ``context''  representing code relevant for $C$. The $n$-gram model, instead, does only consider the $n-1$ preceding tokens to predict the $n^{th}$ token. 

Our results showed the superiority of the T5, achieving significantly better prediction performance as compared to the $n$-gram model. However, the simplicity of the latter and its wider applicability (it does not require a code context as input), make the two models potentially complementary in the implementation of a code comment completion tool.

Our future work aim at perform additional investigations needed to better understand the strengths and weaknesses of the two models. This include studying: (i) the role played by the ``code context'' in the performance of the T5 (\ie what are the performance of the T5 when providing as input \emph{only} the comment tokens ``typed'' by the developer?); (ii) the performance of the experimented models for different types of code comments (\eg Pascarella \etal \cite{Pascarella:emse2019} presented a taxonomy of code comment types and an approach to automatically classify them); (iii) how the T5 performance varies when using larger  models \cite{raffel2019exploring}; and (iv) whether the performance of the models improves when training and testing them on code comments from a specific domain (\eg only comments extracted from Android apps). 

Finally, we are working on the integration of the two models in an IDE plugin to assess their usefulness in studies with developers. 

\section*{Acknowledgment}

This project has received funding from the European Research Council (ERC) under the European Union's Horizon 2020 research and innovation programme (grant agreement No. 851720).

\bibliography{main}
\bibliographystyle{IEEEtran}

\end{document}